# Phonon anharmonicity-driven charge density wave transition and ultrafast dynamics in 1$T$-TaS$_2$/TaSe$_2$

Wenqian Tu,[1,2] Run Lv,[1,2] Dingfu Shao,[1] Yuping Sun,[3,1,4] and Wenjian Lu[1,*]

[1]Key Laboratory of Materials Physics, Institute of Solid State Physics, HFIPS, Chinese Academy of Sciences, Hefei 230031, China
[2]University of Science and Technology of China, Hefei 230026, China
[3]High Magnetic Field Laboratory, HFIPS, Chinese Academy of Sciences, Hefei 230031, China
[4]Collaborative Innovation Center of Microstructures, Nanjing University, Nanjing 210093, China

[*]Corresponding author: wjlu@issp.ac.cn

Charge density wave (CDW), a symmetry-breaking collective phenomenon in condensed matter systems, exhibits periodic modulations of electron density coupled with lattice distortions, where the lattice plays a critical role via electron-phonon coupling. In transition metal dichalcogenides (TMDs) 1$T$-TaS$_2$/TaSe$_2$, experiments reveal rich temperature- and pressure-dependent CDW phase behaviors, along with metastable CDW states induced by ultrafast optical excitation. Nevertheless, the underlying mechanisms governing thermal/pressure-driven transitions and particularly the microscopic evolution of CDW phases remain incompletely understood. Here, we perform first-principles anharmonic phonon calculations and machine-learning force-field molecular dynamics at finite temperatures/pressures to investigate the CDW transitions in 1$T$-TaS$_2$/TaSe$_2$. The calculated CDW transition temperature $T_{CDW}$ and critical pressure $P_c$ are in quantitative agreement with experimental values. Our results demonstrate that the melting of CDW originates from phonon anharmonicity, with ionic fluctuations dominating the transition dynamics. We observe the microscopic evolution of CDW under varying temperature/pressure, revealing an ultrafast nucleation process of CDW (~3 ps). Our results emphasize the essential role of phonon anharmonicity in elucidating CDW transition mechanisms underlying, and advance fundamental understanding of CDW-related phenomena in TMDs.

## I. INTRODUCTION

Charge density wave (CDW) refers to a periodic modulation of the electron density accompanied by a distortion of the crystal lattice, which has been extensively studied in layered transition metal dichalcogenides (TMDs) owing to their unique physical properties [1-3]. Different CDWs have been observed: $1T$-VSe$_2$ exhibits $\sqrt{7}\times\sqrt{3}$ and $4\times4$ CDW superlattices [4], $1T$-TiSe$_2$ displays a $2\times2$ CDW superlattice [5], $1T$-TaS$_2$/TaSe$_2$ adopt $\sqrt{13}\times\sqrt{13}$ CDW superlattices [6], while most TMDs with $2H$ structures are known to exhibit $3\times3$ CDW superlattice [7]. Among the various TMDs, $1T$-TaS$_2$/TaSe$_2$ have attracted significant interest for their exceptionally rich CDW phase diagrams as a function of temperature and pressure: A low-temperature commensurate CDW (CCDW) phase with "star-of-David" (SoD) superstructure evolves through nearly-commensurate (NCCDW, 180K) and incommensurate (ICCDW, 350K) phases before restoring hexagonal symmetry above 550 K in $1T$-TaS$_2$ [8,9]. In contrast, $1T$-TaSe$_2$ undergoes a direct CCDW to ICCDW (473 K) transition without NCCDW intermediates [10]. Hydrostatic pressure can suppress CDW orders, collapsing the CCDW phases at 2.5 GPa ($1T$-TaS$_2$) and 6.5 GPa ($1T$-TaSe$_2$) [11-13], respectively. Moreover, ultrafast optical melting and switching CDW phases have been demonstrated in $1T$-TaS$_2$/TaSe$_2$ [14-18], accessing a range of dynamical pathways and timescales. Despite these advances, fully elucidating the microscopic mechanisms governing thermal and pressure-driven CDW transitions and the microscopic evolution process remains a challenge.

The formation mechanism of the CDW in TMDs was initially attributed to Fermi-surface nesting (FSN) with elastic electronic scatterings [19,20]. Competing theories, however, emphasize the momentum-dependent electron-phonon coupling (EPC) as the primary driving force, where FSN is negligible [21,22]. Harmonic phonon calculations for $1T$-TaS$_2$/TaSe$_2$ can correctly reproduce the occurrence of CDW with $\sqrt{13}\times\sqrt{13}$ periodicity and reveal substantial EPC strength at the soft phonon modes [23,24]. However, the harmonic phonon fundamentally fails to capture finite-temperature lattice dynamics, elucidate CDW transition mechanisms, or explain the high-temperature stability of the $1T$ phase. Recently, non-perturbative studies have demonstrated that phonon anharmonicity, incorporating thermal ionic fluctuations, triggers CDW melting in $2H$-NbS$_2$/NbSe$_2$ [25,26]. Undoubtedly, for an accurate description of the CDW transitions in $1T$-TaS$_2$/TaSe$_2$, phonon calculations incorporating anharmonic effects should be carried out. Furthermore, conventional *ab initio* molecular dynamics (AIMD) cannot access CDW-related phase transition processes due to the prohibitive computational costs of large supercells comprising thousands to tens of thousands of atoms. We overcome this limitation using on-the-fly trained machine-learned force fields (MLFFs) [27-29], enabling large-scale, density functional theory (DFT)-accurate MD simulations at finite temperatures/pressures to investigate the microscopic dynamics of CDW transitions.

In this work, we performed joint first-principles anharmonic phonon calculations and MLFF-AIMD simulations to investigate the CDW transitions in 1$T$-TaS$_2$/TaSe$_2$. Our results reveal that the formation of CDW orders in 1$T$-TaS$_2$/TaSe$_2$ is primarily attributed to the strong EPC. By incorporating phonon anharmonicity into our calculations, we uncover the elimination of harmonic-level structural instabilities and yield accurate predictions of the CDW transition temperature $T_{CDW}$ and critical pressure $P_c$ values. Our results highlight the predominance of ionic fluctuations in melting the CDW phases. The chemical bonding analysis indicates that the relative enhancement of *d-p* hybridization underpins the greater CDW phase stability observed in 1$T$-TaSe$_2$ versus 1$T$-TaS$_2$. Further microscopic dynamic analysis of the CDW transition reveals an ultrafast nucleation process (~3 ps). Our present study not only establishes the dominant role of phonon anharmonicity in driving CDW transitions but also reveals the microscopic evolution process of CDW transition under thermal/pressure.

## II. METHODS

First-principles calculations based on DFT were performed using the Vienna *ab initio* Simulation Package (VASP) [30,31]. The generalized gradient approximation (GGA) in the scheme of Perdew-Burke-Ernzerhof (PBE) [32] was applied for the exchange-correlation function. The kinetic energy cutoff was set to 500 eV. Considering the weak interlayer interactions in bulk, a van der Waals (vdW) correction was included using the nonlocal rev-vdW-DF2 exchange-correlation functional proposed by Hamada [33]. The Brillouin zone (BZ) was segmented by a Gamma-centered grid with a resolution of 0.02 Å$^{-1}$ for self-consistent calculations. For the harmonic phonon dispersion, we calculated harmonic force constants within a 6×6×2 supercell using the finite difference method [34] implemented in PHONOPY [35]. The phonon linewidth and the EPC constant were calculated using the QUANTUM ESPRESSO packages [36].

We trained an MLFF for 1$T$-TaS$_2$/TaSe$_2$ using an on-the-fly method in the DFT calculation process, as implemented in VASP [37]. The initial training datasets were generated via AIMD simulations at 200/300 K using a 6×6×2 supercell to sample thermally perturbed ground-state configurations. AIMD simulations at the DFT level provide precise forces and energies. To obtain the anharmonic phonon spectrum at a finite temperature, the renormalized phonon frequencies were extracted by projecting the MD trajectory onto a set of harmonic phonon modes using the mode decomposition technique as implemented in the DYNAPHOPY code [38]. For the MD simulations, the canonical (*NVT*) ensemble with the Nosé-Hoover thermostat was applied to control the temperature [39].

To investigate the microscopic dynamic process of CDW transitions, we executed MD simulations using the well-trained MLFFs and a quite large 10×10×2 CDW supercell with 7800 atoms. MD simulations were employed in the isobaric-isothermal (*NPT*) ensemble with the temperature set from 5 to 500 K and the pressure from 0 to 6 GPa, in a duration of 30 ps with a timestep of 1 fs. Radial distribution functions (RDFs) were obtained by using the VASPKIT code [40]. The chemical bonding interactions were analyzed using the LOBSTER package [41]. The "pbevaspfit2015" basis set was adopted, including the 5$p$5$d$6$s$ orbitals for Ta atom, 3$s$3$p$ orbitals for S atom, and 4$s$4$p$ orbitals for Se atom.

### III. RESULTS AND DISCUSSION

The crystal structures of 1$T$-TaS$_2$/TaSe$_2$ bulk adopt layered sandwiched structures, with each Ta atom bonded to six S or Se atoms forming octahedral coordination (Fig. 1(a)). Adjacent layers are stabilized by weak vdW interactions. The theoretically optimized lattice constants of 1$T$-TaS$_2$ (3.35 Å) and 1$T$-TaSe$_2$ (3.47 Å) align well with previous experimental reports [42,43]. Among the CDW phases observed in 1$T$-TaS$_2$/TaSe$_2$, the CCDW phase displays the simplest geometric pattern. The CDW transition arises from spontaneous symmetry breaking, accompanied by atomic displacements as illustrated in Fig. 1(b). The low-temperature CCDW phase is characterized by a triplet of CDW wave vectors $\boldsymbol{Q}_{CDW}$ oriented at 120º with respect to each other, yielding a $\sqrt{13}\times\sqrt{13}$ supercell. Within Ta-plane, twelve outer Ta$_1$ and Ta$_2$ atoms shift toward a central Ta$_0$ atom, forming a so-called "SoD" cluster. Given that the displacement of the Ta atoms is much larger than that of S (Se) atoms, we use Ta atomic displacement as an indicator for CDW-type lattice distortion.

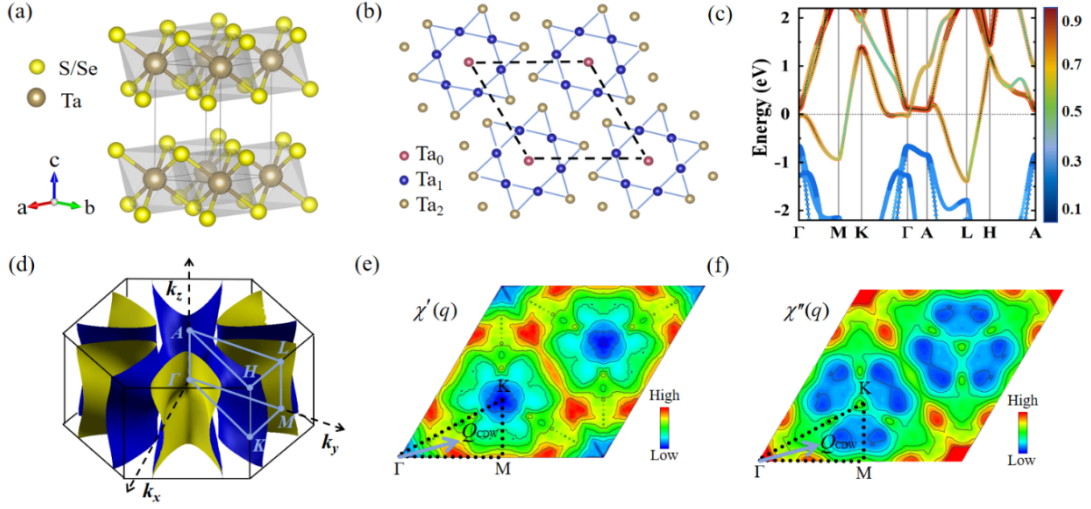

FIG. 1. (a) Crystallographic structure of $1T$-TaS$_2$/TaSe$_2$. (b) Top view of the $\sqrt{13}\times\sqrt{13}$ CCDW phase in the Ta-Ta plane, where the unit cell in the CCDW phase is denoted with black dashed lines. The blue hexagrams are the so-called "SoD" pattern, in which Ta$_1$ and Ta$_2$ atoms are converged towards Ta$_0$ atoms. (c) Projected band structure of $1T$-TaS$_2$, where the contributions of Ta-$5d$ and S-$3p$ orbitals are indicated by red and blue dots, respectively. The brightness of the dots denotes the orbital weights. (d) Fermi surface of $1T$-TaS$_2$. The high-symmetry paths are indicated by blue lines. (e) Real and (f) imaginary parts of the electron susceptibility of $1T$-TaS$_2$ in the $\boldsymbol{Q}_z = 0$ plane, where $\boldsymbol{Q}_{CDW}$ is indicated by the grey arrow.

We first study the electronic band structure of $1T$-TaS$_2$ (Fig. 1(c)) along the high-symmetry $k$-path within the BZ as shown in Fig. 1(d). The calculated bands exhibit a characteristic signature of TMDs in the $1T$ phase, with only one band crossing the Fermi energy level ($E_F$). The orbital-projected band reveals that the electronic states near $E_F$ are predominantly derived from Ta-$5d$ orbitals, underscoring the critical role of Ta atoms in subsequent dynamics investigations. Because of the quasi-two-dimensional (2D) nature of the layered crystal structure, the band around the $E_F$ is almost flat along the $\Gamma$-$A$ direction. The corresponding Fermi surface (FS) shown in Fig. 1(d) corroborates a prior report [44]. A sixfold petal-like FS encircling an electronic pocket at $M$ point exhibits weak dispersion along the $\Gamma$-$A$ direction, further confirming the quasi-2D properties of the electronic state.

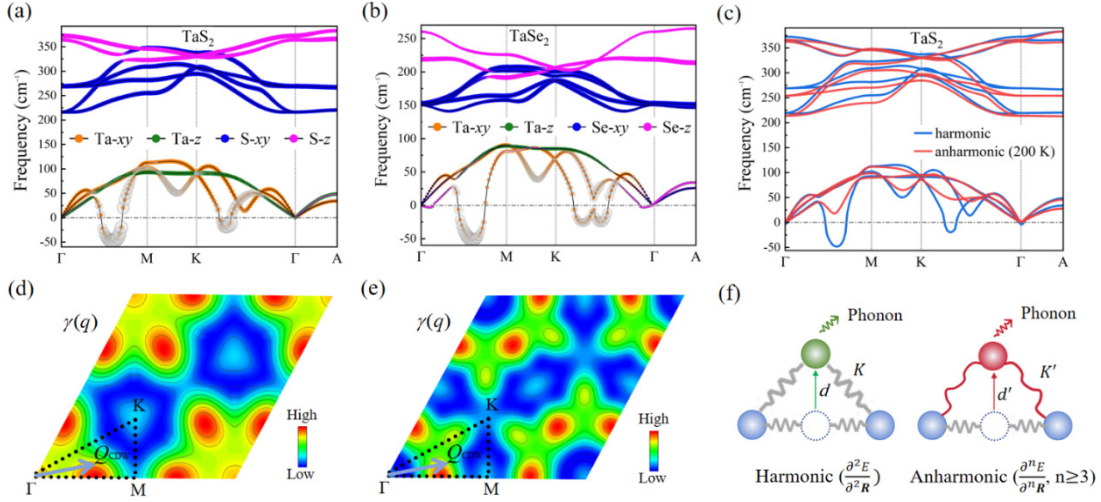

FIG. 2. Harmonic phonon spectra of (a) 1$T$-TaS$_2$ and (b) 1$T$-TaSe$_2$, which are weighted by the vibration modes of Ta and S atoms. The out-of-plane acoustic phonons are decorated with grey symbols, proportional to the partial EPC strength $\lambda_{qv}$. (c) Harmonic versus anharmonic (200K) phonon spectra of 1$T$-TaS$_2$. Phonon linewidth of the lowest phonon mode of (d) 1$T$-TaS$_2$, and (e) 1$T$-TaSe$_2$ in the $Q_z = 0$ plane, where $Q_{CDW}$ is indicated by the grey arrow. (f) Schematic comparison between the harmonic (left panel) and the anharmonic phonon (right panel) scattering processes, represented by simple one-dimensional ball and spring models. The harmonic phonon displaces the atom (green) by an amount $d$ without affecting the spring constant $K$ ($\propto \frac{\partial^2 E}{\partial^2 R}$), while the anharmonic one displaces the atom (red) by a large amount $d'$, inducing a change in the spring constant from $K$ to $K'$ ($\propto \frac{\partial^n E}{\partial^n R}, n \geq 3$).

The formation of CDW order is generally attributed to two theoretical mechanisms: FSN [45] and EPC [23]. FSN attributes CDW formation to elastic electronic scatterings at the FS, particularly in regions where parallel segments of the FS are separated by a single wave vector $Q$. In contrast, EPC emphasizes the interaction between electrons and lattice vibrations through phonon-mediated inelastic scattering processes. For example, CDW in 2$H$-NbSe$_2$ has been explained within the framework of the FSN [46], while EPC has been confirmed as the dominant mechanism for CDW formation in 1$T$-VSe$_2$ [47]. A comprehensive analysis of both FSN and EPC, from electronic and phononic perspectives, is essential for a deeper understanding of CDW formation. To get insight into the origin of the CDW order in 1$T$-TaS$_2$/TaSe$_2$, we calculated both the real ($\chi'$) and imaginary ($\chi''$) parts of electron susceptibility [19,20,48], as well as the phonon linewidth $\gamma$ [22,23]. These parameters are defined by:

$$\chi'(q) = \sum_k \frac{f(\varepsilon_k) - f(\varepsilon_{k+q})}{\varepsilon_k - \varepsilon_{k+q}}, \qquad (1)$$

$$\chi''(q) = \sum_k \delta(\varepsilon_k - \varepsilon_F)\delta(\varepsilon_{k+q} - \varepsilon_F), \qquad (2)$$

$$\gamma_{qv} = 2\pi\omega_{qv} \sum_{ij} \int \frac{d^3k}{\Omega_{BZ}} |g_{qv}(k,i,j)|^2 \times \delta(\varepsilon_{q,i} - \varepsilon_F)\delta(\varepsilon_{k+q,i} - \varepsilon_F), \qquad (3)$$

$$g_{qv}(k,i,j) = \left(\frac{\hbar}{2M\omega_{qv}}\right)^{1/2} \left\langle \psi_{i,k} \left| \frac{dV_{SCF}}{d\hat{u}_{qv}} \hat{\xi}_{qv} \right| \psi_{j,k+q} \right\rangle. \qquad (4)$$

Here $f(\varepsilon_k)$ is the Fermi-Dirac function, $\varepsilon_k$ is the corresponding Kohn-Sham energy, $\varepsilon_F$ is the Fermi energy level, $\psi$ is the wave function, $V_{SCF}$ is the Kohn-Sham potential, $\hat{u}$ is atomic displacement, and $\hat{\xi}$ is the phonon eigenvector. Figures 1(e) and (f) show the $\chi'$ and $\chi''$ of the electron susceptibility for 1$T$-TaS$_2$ in the $\boldsymbol{Q}_z = 0$ plane, respectively. The real part $\chi'$ governs the stability of the electronic system, while the imaginary part $\chi''$ represents the low-frequency limit of the bare electronic susceptibility. Therefore, if the CDW order is induced by FSN, both the $\chi'$ and $\chi''$ should exhibit peaks at the $\boldsymbol{Q}_{\text{CDW}} = \frac{3}{13}\mathbf{a}^* + \frac{1}{13}\mathbf{b}^*$. However, our calculations show that the maxima of $\chi'$ and $\chi''$ are displaced from the $\boldsymbol{Q}_{\text{CDW}}$, indicating that FSN does not play a dominant role in CDW formation for 1$T$-TaS$_2$. Due to the striking similarity in structural and electronic properties between 1$T$-TaS$_2$ and 1$T$-TaSe$_2$, the representative results presented here focus on 1$T$-TaS$_2$ to avoid redundancy while maintaining generalizable insights. Conversely, the phonon linewidth $\gamma$ displays sharp peaks around $\boldsymbol{Q}_{\text{CDW}} = \frac{3}{13}\mathbf{a}^* + \frac{1}{13}\mathbf{b}^*$ for both 1$T$-TaS$_2$ and 1$T$-TaSe$_2$, as shown in Figs. 2(d) and (e). This observation underscores the strong momentum dependence of the electron-phonon matrix elements, pointing to EPC as the primary mechanism driving CDW orders. Our present findings are consistent with previous works [23,24], thereby highlighting the critical role of the lattice via EPC in accurately describing the CDW properties of 1$T$-TaS$_2$/TaSe$_2$.

Harmonic phonon spectrum calculation has been proven to be an effective method for modeling CDW instabilities, as the emergence of imaginary phonon frequencies for high-symmetry structures directly signals lattice reconfiguration toward superlattice ordering. Usually, dynamical instability is evidenced by imaginary frequencies at $Q_{CDW}$, signaling superlattice formation with a commensurate wavevector [49]. Our harmonic phonon spectra for $1T$-TaS$_2$/TaSe$_2$ (Figs. 2(a) and (b)) reveal imaginary frequencies in acoustic modes, clearly indicating structural instability. The phonon instability near $\frac{1}{2}\Gamma M$ is related to the $\sqrt{13}\times\sqrt{13}$ CDW modulation. The imaginary modes originate mostly from the in-plane vibration of Ta atoms (Ta-$xy$), complemented by minor contributions from the out-of-plane vibration of S or Se atoms (S-$z$ or Se-$z$). The higher atomic mass of Se compared to S results in a larger contribution to acoustic phonon branches. The softening of acoustic modes is closely tied to the bonding between Ta and its chalcogen neighbors (S/Se). Such bonding is possible to be anharmonic and may induce the CDW transition upon heating, pressure, or doping/defect. The strength of EPC can be estimated by calculating the EPC constant $\lambda(\omega)$ using the formula

$$\lambda(\omega)=\sum_{qv}\lambda_{qv}=2\int\frac{\alpha^2 F(\omega)}{\omega}d\omega, \tag{5}$$

where $\omega$ is the phonon frequency, $\lambda_{qv}$ is the EPC constant contributed by the vth mode at the wave vector $q$, and $\alpha^2 F(\omega)$ is the Eliashberg spectral function, defined as:

$$\alpha^2 F(\omega)=\frac{1}{2\pi N(E_F)}\sum_{qv}\delta(\omega-\omega_{qv})\frac{\gamma_{qv}}{\hbar\omega_{qv}}. \tag{6}$$

We decorate the magnitude of $\lambda_{qv}$ in the out-of-plane acoustic phonon branch (usually referred to as the ZA mode) of phonon dispersion with grey symbols, which is estimated by

$$\lambda_{qv}=\frac{\gamma_{qv}}{\pi\hbar N(E_F)\omega_{qv}^2}. \tag{7}$$

According to the above definition, phonon modes with a lower frequency will lead to stronger EPC. Therefore, soft acoustic modes around the $Q_{CDW}$ significantly contribute to the total EPC, as shown in Figs. 2(a) and (b).

Figure 2(c) shows the comparison of harmonic and anharmonic (200 K) phonon spectra of $1T$-TaS$_2$. When the temperature increases to 200 K, the imaginary frequencies around the $Q_{CDW}$ vector vanish, consistent with the experimental observation of $T_{CDW}$ ≈ 180 K. The phonon harmonic interaction considers only the second-order derivative of the lattice potential, while anharmonic interactions involve third- and higher-order derivatives. The higher-order terms enable coupling between distinct phonon modes, thereby introducing temperature-dependent effects into the system. Conceptually visualizing interatomic forces via ball-and-spring models (Fig. 2(f)) elucidate this distinction: Harmonic phonon displaces atom (green) by an amount $d$ without affecting the spring constant $K$ ($\propto \frac{\partial^2 E}{\partial^2 R}$), while the anharmonic one displaces atom (red) by a large amount $d'$ inducing a change in the spring constant from $K$ to $K'$ ($\propto \frac{\partial^n E}{\partial^n R}, n \geq 3$).

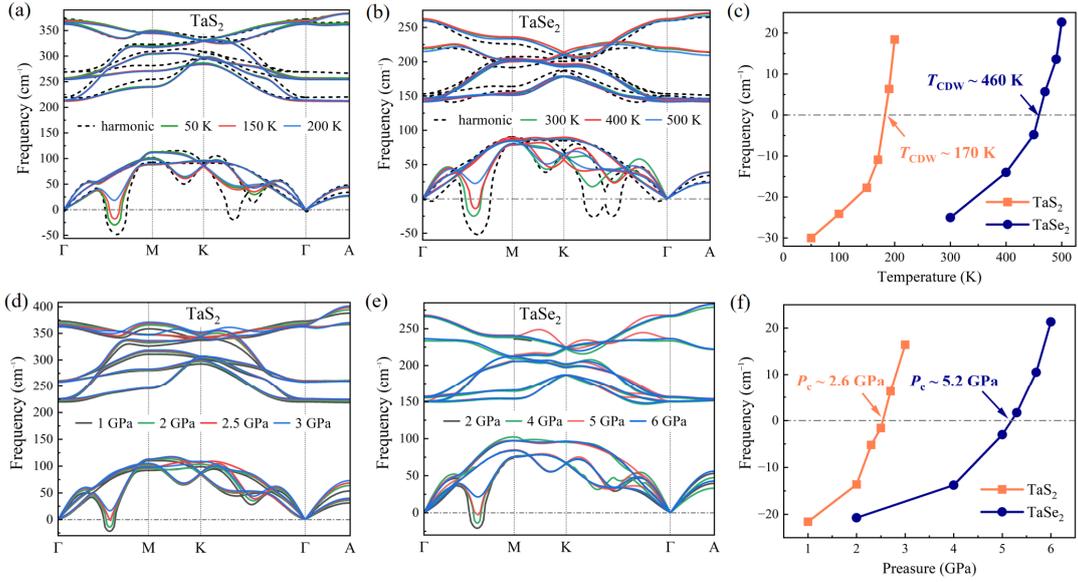

FIG. 3. Harmonic (black dashed lines) and anharmonic phonon spectra at different temperatures (colored solid lines) of (a) 1$T$-TaS$_2$ and (b) 1$T$-TaSe$_2$. (c) Temperature dependence of the out-of-plane acoustic mode frequency at $\mathbf{Q}_{CDW} \approx \frac{1}{2}\Gamma M$. The $T_{CDW}$ is evaluated at 170 K for 1$T$-TaS$_2$ (orange line) and 460 K for 1$T$-TaSe$_2$ (blue line). Pressure-dependent anharmonic phonon spectra of (d) 1$T$-TaS$_2$ and (e) 1$T$-TaSe$_2$ at 10 K. (f) Pressure dependence of the out-of-plane acoustic mode frequency at $\mathbf{Q}_{CDW} \approx \frac{1}{2}\Gamma M$. The $P_c$ is estimated at 2.6 GPa for 1$T$-TaS$_2$ (orange line) and 5.2 GPa for 1$T$-TaSe$_2$ (blue line).

To provide a more comprehensive understanding, we present the temperature-dependent anharmonic phonon spectra of 1$T$-TaS$_2$/TaSe$_2$ at various temperatures in Figs. 3(a) and (b). It is evident that phonon anharmonicity significantly modulates the soft modes, even in the low-temperature regime. As the temperature increases, the imaginary frequencies around $\mathbf{Q}_{CDW}$ gradually diminish, suggesting that phonon anharmonic interactions suppress the CDW instabilities. One can also find the rapid dissipation of instability around $\boldsymbol{Q} \approx \frac{2}{3}\Gamma K$ even at low temperatures. This highlights the inherent limitations of phonon harmonic approximations, which erroneously predict phonon instability regions. Judging by the disappearance of soft modes at $\boldsymbol{Q}_{CDW} \approx \frac{1}{2}\Gamma M$ from the anharmonic phonon spectra, we can determine the $T_{CDW}$ to be approximately 170 K for 1$T$-TaS$_2$ and 460 K for 1$T$-TaSe$_2$ (see Fig. 3(c)). The theoretically obtained $T_{CDW}$ values from anharmonic phonon spectra show a good agreement with experimental values (180 K for 1$T$-TaS$_2$ and 473 K for 1$T$-TaSe$_2$).

We further investigate the pressure-induced CDW transition in 1$T$-TaS$_2$/TaSe$_2$. In Figs. 3(d) and (e), we show the anharmonic phonon spectra of 1$T$-TaS$_2$/TaSe$_2$ at 10 K under various pressures. The results reveal that the unstable softening modes associated with CDW formation in 1$T$-TaS$_2$/TaSe$_2$ exhibit notable hardening under pressure. We can evaluate the critical pressure $P_c$, at which the CDWs are fully suppressed. The theoretically obtained $P_c$ values for 1$T$-TaS$_2$ and 1$T$-TaSe$_2$ are 2.6 GPa and 5.2 GPa, respectively (Fig. 3(f)). The theoretical $P_c$ values deduced from the anharmonic phonon spectra show a good agreement with experimental observations (2.5 GPa for 1$T$-TaS$_2$ and 6.5 GPa for 1$T$-TaSe$_2$). Our findings underscore the importance of incorporating phonon anharmonic effects for a more realistic description of pressure-induced CDW phase transitions.

One can note that 1$T$-TaSe$_2$ has higher $T_{CDW}$ and $P_c$ compared with 1$T$-TaS$_2$. It may be attributed to the stronger atomic interactions in 1$T$-TaSe$_2$, which stabilize the CDW structure. To elucidate the differences in chemical bonding of 1$T$-TaS$_2$/TaSe$_2$, we made a comparative analysis of interatomic bonding and charge transfer in 1$T$-TaS$_2$/TaSe$_2$. We first calculate the crystal orbital Hamilton populations (COHP) [50] of 1$T$-TaS$_2$ and 1$T$-TaSe$_2$, revealing the nature of occupied and unoccupied states. The strength of the bonds can be evaluated using the integrated COHP (ICOHP) values, which quantitatively describe the $d$-$p$ or $d$-$d$ orbital hybridization. A more negative ICOHP value corresponds to a higher degree of orbital hybridization. As depicted in Figs. 4(a), (c), (e), and (g), the ICOHP of Ta-S/Se bonds are much smaller compared to those of Ta-Ta bonds, indicating that Ta-S/Se dominates in chemical bond interactions. In the CDW phase (SoD superstructure), the ICOHP values of the Ta-S/Se bonds demonstrate a significant decrease, indicating a substantial enhancement of the $d$-$p$ hybridization of the Ta-S/Se bonds upon CDW structural transition. With CDW formation, the ICOHP values of Ta-S/Se bonds decrease by 9.2% (-3.34 to -3.65) for TaS$_2$ and 17.2% (-3.02 to -3.54) for TaSe$_2$, respectively. These results may indicate a significant enhancement of *d-p* hybridization in the CDW state of 1*T*-TaSe$_2$ compared to 1*T*-TaS$_2$. We also calculate the differential charge density to visualize the tendencies of the charge redistribution, as shown in Figs. 4(b), (d), (f), and (h). The results demonstrate the electron depletion at Ta atom sites and the corresponding electron accumulation at S/Se atom sites. Compared with the ideal 1*T* phase (Figs. 4(b) and (f)), the CDW phase (Figs. 4(d) and (h)) exhibits a more pronounced charge polarization: Significant electron depletion at Ta atoms coincides with enhanced electron accumulation at S/Se atoms. The above results imply that the enhanced *d-p* hybridization facilitates the stabilization of CDW phases in both 1*T*-TaS$_2$ and 1*T*-TaSe$_2$.

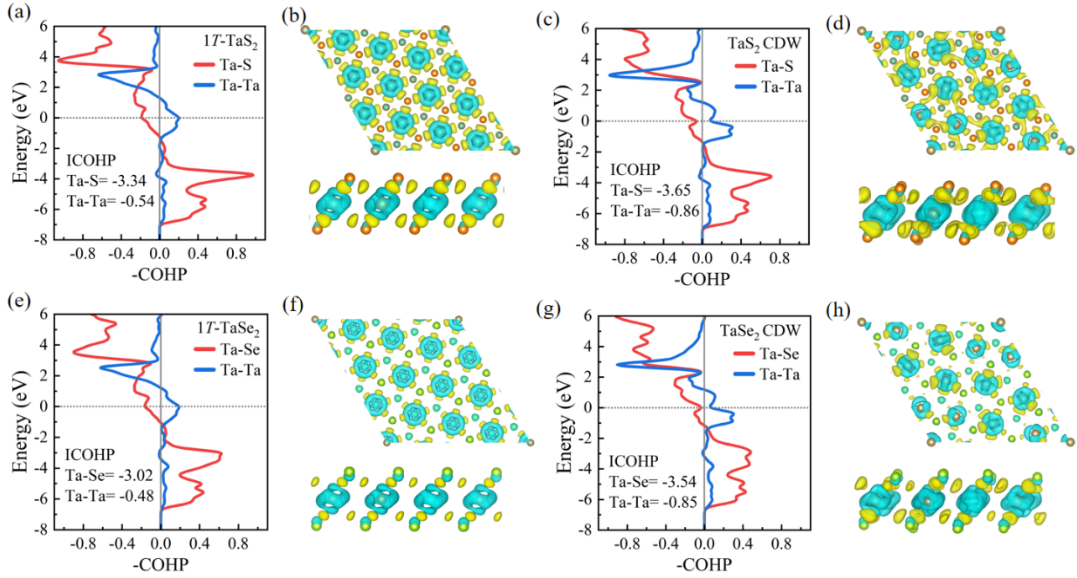

FIG. 4. COHP and the corresponding averaged ICOHP values of Ta-Ta and Ta-S/Se bonds for (a) ideal 1*T* phase, (c) $\sqrt{13}\times\sqrt{13}$ CDW phase of TaS$_2$, (e) ideal 1*T* phase, and (g) $\sqrt{13}\times\sqrt{13}$ CDW phase of TaSe$_2$. (b), (d), (f), and (h) Side- and top-views of differential charge density isosurfaces for TaS$_2$ and TaSe$_2$ in the ideal 1*T* phase and the $\sqrt{13}\times\sqrt{13}$ CDW phase. Yellow regions represent the charge accumulation, and blue regions represent the charge consumption.

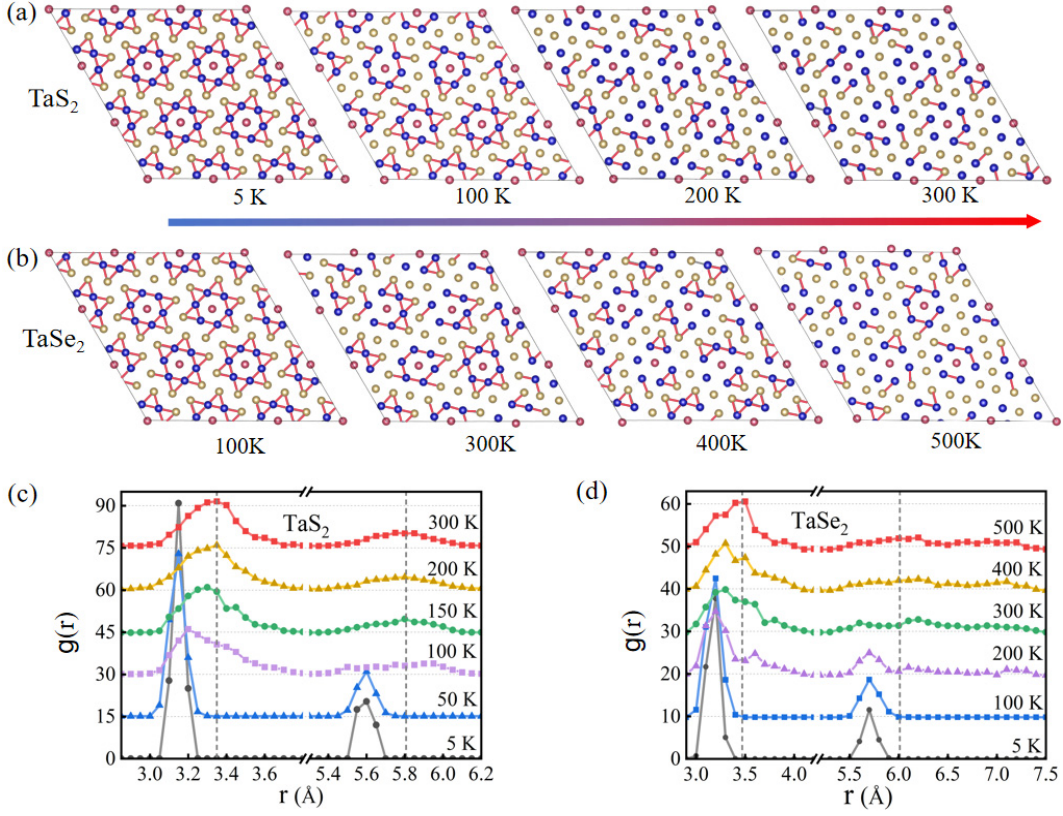

FIG. 5. Temperature-dependent microscopic structural evolutions of CDW transitions in the Ta-Ta plane of (a) $1T$-TaS$_2$ and (b) $1T$-TaSe$_2$ revealed by MLFF-AIMD simulations. The red, blue, and yellow balls represent Ta$_0$, Ta$_1$, and Ta$_2$ atoms, respectively. For the sake of clarity, only partial regions of the simulated $10\times10\times2$ CDW supercell at several temperatures are shown. (c) and (d) RDF corresponding to the simulated temperature-dependent supercell data. The dotted vertical lines indicate the equilibrium nearest-neighbor (Ta$_0$ and Ta$_1$) and next-nearest-neighbor (Ta$_0$ and Ta$_2$) bond distances of the ideal $1T$ phase at high temperature. For clarity, data at different temperatures in (c) and (d) are vertically offset by a fixed constant.

To investigate the underlying microscopic process of the CDW phase transitions in $1T$-TaS$_2$/TaSe$_2$, we perform the MLFF-AIMD simulations at finite temperatures/pressures. Here, we focus on tracking Ta atomic configurations to elucidate the atomistic pathways of structural reorganization during CDW phase evolution. As shown in Figs. 5(a) and (b), heating processes induce the progressive transformations from ordered CCDW superstructures to disordered CDW structures for TaS$_2$ and TaSe$_2$ supercells. Initially, we set ~3% displacement of Ta atoms constructing the CCDW superstructures to obtain a randomly perturbed configuration at 5 K.

Subsequent heating reveals gradual dissolution of "SoD" clusters and recovery toward high-temperature $1T$ structures. Quantitative analysis of the phase transition dynamics requires defining order parameters sensitive to structural transformations. Here, we use RDF as a primary metric for analyzing the nearest-neighbor ($Ta_0$ and $Ta_1$) and the next-nearest-neighbor ($Ta_0$ and $Ta_2$) distances between Ta atoms. The temperature-dependent RDF evolution in Figs. 5(c) and (d) provide critical insights: In the low-temperature CCDW phase, cooperative displacements of $Ta_1$ and $Ta_2$ atoms toward $Ta_0$ sites induce dual peaks in the Ta-Ta RDF at nearest-neighbor and next-nearest-neighbor positions. Upon heating, progressive suppression of the CCDW superstructure drives $Ta_1$ and $Ta_2$ atoms away from $Ta_0$, causing RDF peak shifts toward the high-symmetry positions characteristic of the $1T$-phase.

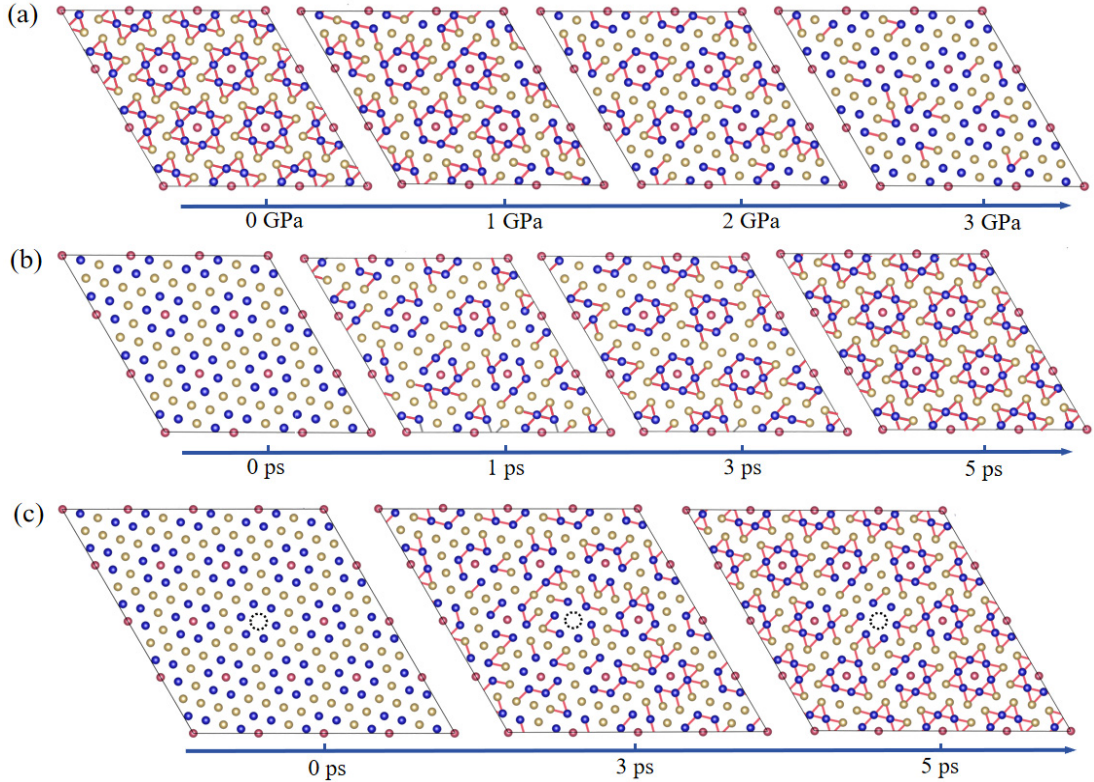

FIG. 6. (a) Pressure- and (b) time-dependent microscopic evolutions of the CDW phase of $1T$-$TaS_2$. The MD simulating temperature is set to 5 K. (c) Ta-vacancy effect on microscopic evolutions of CDW phase. The red, blue, and yellow balls represent $Ta_0$, $Ta_1$, and $Ta_2$ atoms, respectively. The black dashed circle presents the Ta vacancy.

We further investigate the microscopic dynamical processes of the CDW phase in $1T$-$TaS_2$ under pressure. In Fig. 6(a), we simulate the CCDW superstructure under pressure. With increasing pressure, the CDW-related ordered "SoD" clusters gradually disintegrate until reaching ~3 GPa, above which the CDW phase transforms into a normal phase. In Fig. 6(b), we show the CDW nucleation process by MD simulation under ambient pressure at 5 K. One can find that CDW-like lattice distortions emerge in localized regions after 1 ps, followed by CDW nucleation at ~3 ps. Subsequently, the full CDW-related "SoD" clusters fully form at 5 ps. The above results suggest that the CDW formation has an ultrafast nucleation process. Previous time-resolved optical excitation experiments showed that laser irradiation induces a CDW transition on the picosecond time scale without thermal activation [16,51,52], which is consistent with our computed fast recovery of the CDW in ~3 ps. In Fig. 6(c), we also simulate the CDW formation in the presence of Ta atom vacancies. Compared to the defect-free system (Fig. 6(b)), "SoD" clusters progressively form in regions distant from the Ta vacancy. CDW ordering near the Ta vacancy is significantly suppressed despite intact cluster development elsewhere. Our simulation results are consistent with the experimental findings that defects hinder CDW formation [53]. In summary, our results provide a comprehensive understanding of how thermal, pressure, and atomic defects modulate the stability and transitional pathways of CDW phases. Such understanding contributes to a broader comprehension of CDW transitions in TMDs, offering critical guidance for material engineering.

## IV. CONCLUSION

In conclusion, our study employed a synergistic computational framework integrating DFT and MLFF-AIMD, which systematically elucidates the underlying mechanism governing CDW transitions in $1T$-TaS$_2$/TaSe$_2$. Our results identify the strong momentum-dependent EPC as the primary driving force behind the CDW formation. Especially, the anharmonic phonon spectra can accurately capture the dominant role of thermal ionic fluctuations in the CDW melting. Microscopic dynamic analysis of large-scale MD simulations resolves ultrafast dynamics of CDW nucleation, and also demonstrates how thermal, pressure, and defects govern CDW transition. Our study not only deepens the understanding of the complex interplay between electron and lattice in $1T$-TaS$_2$/TaSe$_2$, but also provides theoretical guidance for predicting and manipulating CDW-related phenomena in TMDs.

**ACKNOWLEDGEMENTS**

This work was supported by the National Key Research and Development Program of China under Contract No. 2022YFA1403203. The calculations were performed at Hefei Advanced Computing Center.